\documentclass[mnras,twocolumn,twocolappendix,numberedappendix]{openjournal}
\usepackage{amsmath}
\usepackage{booktabs}
\usepackage{multirow}
\usepackage{color}
\usepackage{soul}
\usepackage{threeparttable}
\usepackage{float}
\usepackage{graphicx}
\usepackage{CJK}
\usepackage{xspace}
\usepackage{afterpage}
\usepackage{placeins}
\usepackage{amssymb}
\usepackage{bm}
 \usepackage{lastpage}
\usepackage[breaklinks,colorlinks,citecolor=blue,urlcolor=blue,linkcolor=blue,filecolor=blue]{hyperref}
\usepackage[normalem]{ulem}

\newcommand{\orcidauthor}[3]{\author{\href{http://orcid.org/#1}{#2$^{#3}$}}}
\newcommand{\Myr}{~\text{Myr}}

\newcommand{\Msun}{~\text{M}_\odot}
\newcommand{\pc}{~\text{pc}}
\newcommand{\kpc}{~\text{kpc}}

\newcommand{\cMpc}{~\text{cMpc}}

\newcommand{\percent}{~\text{per~cent}}
\newcommand{\AAA}{{\text{\AA}}}
\newcommand{\bh}{{\rm BH}}

\newcommand{\effrad}{{\epsilon_{\rm r}}}
\newcommand{\deltamag}{{\Delta\mathrm{mag}}}

\newcommand{\simname}{\textsc{meli$\odot$ra}\xspace}
\newcommand{\ramses}{\textsc{ramses}\xspace}

\newcommand{\boxsizeL}{{10}} 
\newcommand{\finalz}{{8.5}} 
\newcommand{\levelmin}{{9}} 
\newcommand{\levelmax}{{19}} 
\newcommand{\spatialresmin}{{19}} 
\newcommand{\spatialresmax}{{20}} 
\newcommand{\nsf}{{10}} 
\newcommand{\mstarres}{{10^2}} 

\definecolor{mycolor_1}{RGB}{220,30,155}

\newcommand{\NEW}[1]{#1}

\shorttitle{UV diversity of LRDs and DCBH ages}
\shortauthors{Cenci et al.}

\usepackage[export]{adjustbox}

\begin{document}

\title{\vspace{-.5cm}Ultraviolet diversity of Little Red Dots as a probe\\ for direct-collapse black hole ages\vspace{-1cm}}

\orcidauthor{0000-0002-0766-1704}{Elia Cenci}{\dagger}
\orcidauthor{0000-0003-4750-0187}{Melanie Habouzit}{\dagger}
\orcidauthor{0000-0002-8360-3880}{Dale D. Kocevski}{\ddag}

\affiliation{$^\dagger$ Department of Astronomy, University of Geneva, Chemin Pegasi 51, Versoix CH-1290, Switzerland}
\affiliation{$^\ddag$ Department of Physics and Astronomy, Colby College, Waterville, 04901, ME, USA}

\thanks{E-mail:\href{mailto:elia.cenci@unige.ch}{elia.cenci@unige.ch}}

\begin{abstract}
    Little Red Dots (LRDs) uncovered by the James Webb Space Telescope have been proposed as candidate galaxies hosting embedded accreting direct-collapse black holes (DCBHs), yet the relative ultraviolet (UV) emission of their host galaxy remains highly uncertain and diverse across the population. Using a large-scale cosmological hydrodynamical simulation from the \textsc{meli$\odot$ra} suite, we investigate the contribution of PopIII stars and accreting DCBHs in LRD candidates at $z>8.5$, in the rest-frame $0.2-0.6~\mu\mathrm{m}$ band. We find that the UV emission from the host galaxy evolves rapidly over the first $\sim 30~\mathrm{Myr}$ following DCBH formation, reflecting the build-up of stellar mass and metal enrichment. This evolution consists of a rapid transition from initially BH-dominated systems, with negligible stellar mass, low metallicity, and high accretion rates, to progressively more developed hosts in which rapid star formation enhances the UV output and metallicity increases. After $\sim 30~\mathrm{Myr}$, the stellar continuum typically overwhelms the accreting DCBH contribution, producing bluer colours and more extended stellar distributions. As a result, UV-bright LRDs are predicted to host older DCBHs, have higher gas-phase metallicities, lower BH-to-stellar mass ratios, and lower Eddington ratios. The short-lived nature of the LRD phase places strong constraints on their emergence over cosmic time. Overall, our results suggest that DCBH ages can be constrained from the host galaxy contribution to the UV-optical spectrum of LRDs, relative to that of the accreting DCBH, and support the picture in which a DCBH evolutionary sequence is systematically encoded in emission line properties, gas-phase metallicities, and accretion states.
\end{abstract}

\section{Introduction}\label{sec:introduction}

\begin{figure*}
    \centering
    \includegraphics[width=0.8\hsize]{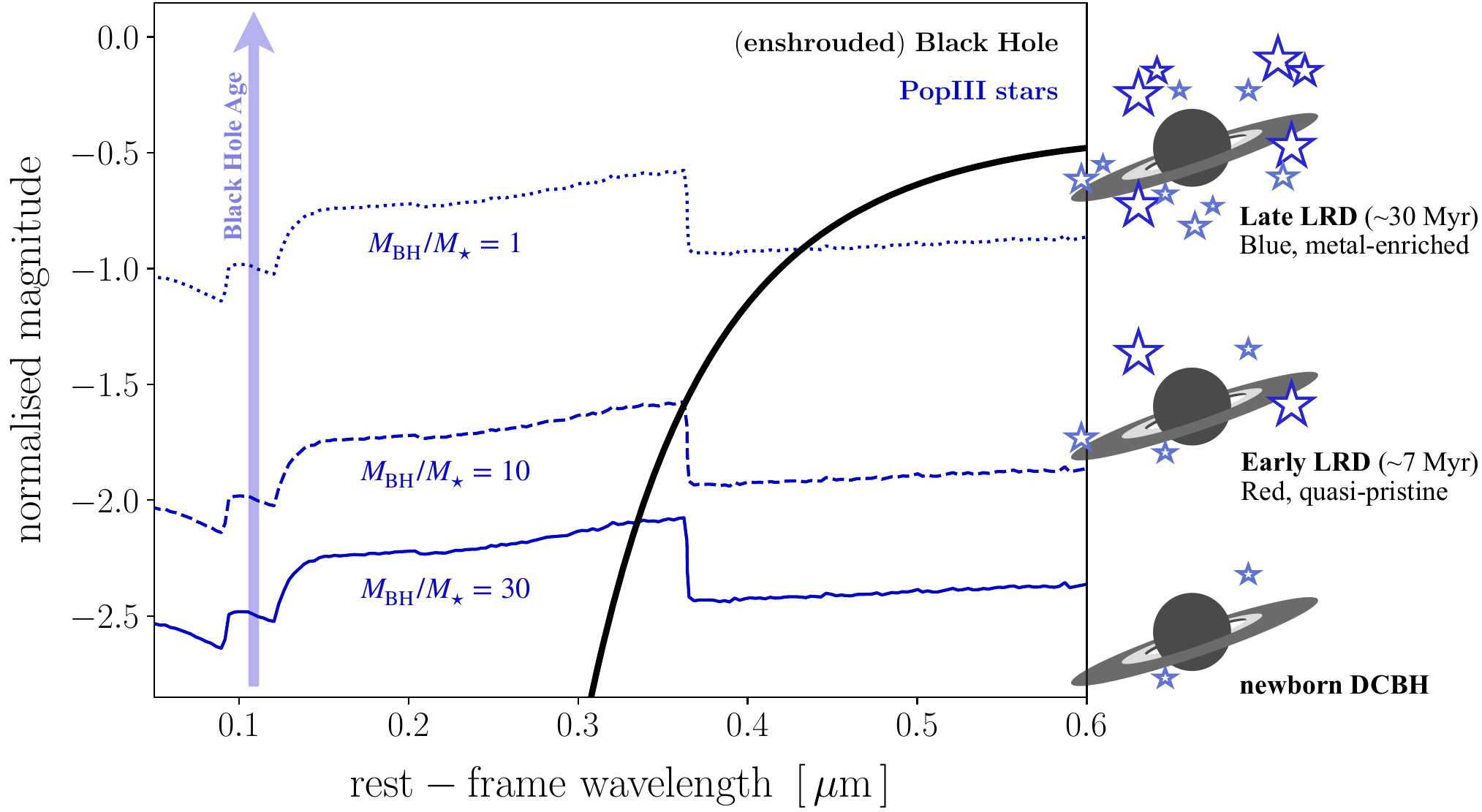}
    \caption{Schematic illustration of the evolutionary sequence of LRD candidate DCBHs. We show idealized rest-frame UV-optical composite SEDs of PopIII stars (blue curves) and an enshrouded accreting DCBH (black curve), for different BH-to-stellar mass ratios. Immediately after DCBH formation \NEW{(early LRD phase; $\mathrm{DCBH~age}<10\,\mathrm{Myr}$, see Section~\ref{sec:results})}, the stellar mass is negligible and the UV continuum is weak, yielding BH-dominated emission and red UV-optical colours. As the system evolves over the first $\sim 30\Myr$ \NEW{(late LRD phase; $\mathrm{DCBH~age}=10-30\,\mathrm{Myr}$)}, rapid star formation increases the stellar UV output, progressively reducing the host-to-core contrast. At later times ($>30\Myr$; post-LRD phase), stellar emission dominates the UV-optical SED and the system exits LRD-like colour selection, also exhibiting an extended ($R_\star>100\pc$) stellar mass distribution. The monotonic evolution of the UV-optical balance provides a physically motivated age sequence for high-$z$ LRD candidates and links their observed diversity to different stages of early DCBH growth.}
    \label{fig:SED_cartoon}
\end{figure*}

The James Webb Space Telescope (JWST) has uncovered an abundant population of compact, optically red sources dubbed `Little Red Dots' \citep[LRDs;][]{Matthee2024a}, in the first billion years of cosmic history ($z\sim2-11$). LRDs are characterised by red rest-optical colours and comparatively blue rest-ultraviolet (UV) slopes, with a `V-shaped' UV-optical continuum inflection around the Balmer limit. Furthermore, LRDs exhibit broad Balmer emission \citep[with $\mathrm{FWHM}\gtrsim 2000~\mathrm{km/s}$;][]{Greene2024,Barro2024,Kocevski2025,Taylor2025a}, likely tied to faint broad-line active galactic nuclei (AGN) powered by accreting massive black holes (BHs). LRDs are relatively ubiquitous in the early Universe, with a population that outnumbers that of $z<9$ quasars by an order of magnitude, to then swiftly decline below $z\lesssim 6$ \citep[e.g.,][]{Kocevski2025,Inayoshi2025,Cenci_and_Habouzit2025}. To explain their peculiar spectra, commonly invoked scenarios involve either dusty star formation \citep[e.g.,][]{Baggen2024,Barro2024,Williams2024} or a reddened AGN \citep[e.g.,][]{Onoue2023,Kocevski2023,Labbe2024,Ubler2024,Li_Z_2025,Volonteri2025}. Alternative explanations for the origin of LRDs include BH-stars \citep[][]{Naidu2025}, quasi-stars \citep[][]{Begelman_and_Dexter2026,Roman-Garza2026}, and self-gravitating disks accreting around supermassive stars \citep[][]{Zwick2025}. However, LRDs may not trace a homogeneous population sharing a common physical origin and appearance, possibly due to varied selection criteria \citep[e.g.][]{Hviding2025}.

Multi-wavelength studies reveal that LRDs are intrinsically X-ray weak \citep[e.g.,][]{Ananna2024,Yue2024,Tortosa2026,Rusakov2026}, suggesting either extremely high hydrogen column densities \citep[e.g.,][]{Maiolino2025a} or BHs accreting above their Eddington limit \citep[e.g.,][]{Inayoshi2024,Lambrides2024,Pacucci_and_Narayan2024,Madau2025}. Furthermore, LRDs have a clear deficit of both hot and cold dust, disfavouring dust-obscured starbursts as the dominant power source \citep[e.g.,][]{Perez-Gonzalez2024,Setton2025}. Recent studies argue that the characteristic turnover around Balmer break wavelengths in LRDs is physically linked to $n=2$ hydrogen levels rather than to an arbitrary intersection of smooth, unattenuated components \citep[][]{Inayoshi_and_Maiolino2025}. Individual LRDs further show that dense gas absorption around a strong ionising source can mimic a stellar Balmer break \citep{Naidu2025,de_Graaff2025,Kokorev2025,Taylor2025b}.

In general, the relative BH and stellar contributions in LRDs remain uncertain and can vary strongly across the population. Virial methods applied to broad Balmer lines yield BH masses $M_{\bh}\sim 10^{7-9}\Msun$ and moderate Eddington ratios in some datasets \citep[e.g.,][]{Greene2024}, while X-ray stacking and population-level spectral analyses favour lower BH masses, $M_{\bh}\sim10^{6-7}\Msun$, hosted by $\sim10^{8-9}\Msun$ galaxies and accreting at higher Eddington ratios \citep[e.g.,][]{Ananna2024,Naidu2025,Sun2026}. Crucially, radiative transfer through dense gas can systematically bias standard AGN estimators, with electron scattering and non-virial line broadening implying BH masses well below values derived with virial methods \citep[e.g.,][]{Inayoshi_and_Maiolino2025,Rusakov2026}. Furthermore, updated corrections on bolometric luminosities may further reduce the inferred contribution from an accreting BH \citep[e.g.,][]{Greene2026}. The possible contribution of reprocessed light to the rest-frame optical continuum, and the relation between stellar and dynamical masses can affect the inferred $M_{\bh}/M_\star$ ratio \citep[e.g.,][]{Maiolino2026,de_Graaff2025,Greene2026}.

Most likely, LRDs represent a heterogeneous population, with substantial diversity in UV slopes, line ratios, and inferred host--core decompositions. The rest-frame UV and optical continua in LRDs may trace two physically distinct emission channels: a host component, associated with stellar emission and dominating the UV, and a compact, reprocessed AGN component dominating the rest-frame optical. Variations in the relative contribution of these two components can generate a sequence of diverse spectra. Recent spectroscopic and panchromatic studies support this picture. \NEW{\citet{Barro2026b} use a sample of JWST-NIRSpec spectra find that LRDs follow a correlated sequence in UV slope, optical-continuum shape, Balmer-break strength, and emission-line diagnostics, which can be reproduced by varying the luminosity and obscuration of a gas-enshrouded AGN relative to the host galaxy.} Furthermore, using JWST-NIRSpec spectra and available MIRI photometry, \citet{Perez-Gonzalez2026} argue that LRDs are spectroscopically heterogeneous and that their UV--optical diversity can be organized primarily as a changing host contribution rather than a single uniform physical class. A similar picture emerges from the diversity of line profiles. \citet{Matthee2026} find that the H$\alpha$ profiles vary systematically with UV--optical colour and strength of the Balmer break: bluer sources have cores with relatively narrow lines, while redder sources show blue-shifted absorption. Such trends suggest that continuum diversity is coupled to changes in gas column density, covering factor, and outflow geometry, rather than being solely a consequence of different stellar population ages or dust reddening. Reported trends between Balmer-break strength, absorption features, and line-profile diversity are qualitatively consistent with this picture \citep[e.g.,][]{de_Graaff2025}, with increasing [OIII] strength marking the emergence of a more developed, metal-enriched, star-forming host around the AGN.

The physical origin of the first BHs remains a central open problem for modern astrophysics. \NEW{Massive BH seed formation channels are broadly divided by their characteristic mass and formation pathway. Light seeds, with $M_{\bh}\sim 10^{2-3}\Msun$, are thought to be remnants of the first, massive, metal-free (PopIII) stars \citep[e.g.,][]{Fryer2001,Madau_and_Rees2001,Xu2013,Smith2018}. Heavy seeds, with $M_{\bh}\sim 10^{4-6}\Msun$, instead form through the isothermal and monolithic collapse of pristine gas in atomic-cooling haloes \citep[e.g.,][]{Loeb_and_Rasio1994,Eisenstein_and_Loeb1995,Haiman1996a,Koushiappas2004,Begelman2006,Lodato_and_Natarajan2006,Latif2013a,Johnson2014,Choi2015}, that undergoes an intermediate supermassive star (SMS) phase with substantial accretion rates \citep[$\sim 0.01-1\Msun\,\mathrm{yr}^{-1}$; e.g.,][]{Begelman2008,Begelman2010,Latif2013e,Prieto2013,Matsukoba2019,Nandal2023,Prole2024a}, and finally results in a direct-collapse black hole (DCBH).} Heavy seeds channels, such as DCBHs, provide a plausible theoretical route to rapid early growth, but their predicted observational signatures depend largely on the coupling between accretion physics and the evolving properties of their host galaxy. Major uncertainties in DCBH modelling stem from the stringent environmental requirements for their formation. \NEW{These include a nearly pristine gas supply \citep[with metallicities $Z/Z_\odot \lesssim 10^{-5} - 10^{-4}$; e.g.,][]{Bromm_and_Loeb2003,Clark2008,Omukai2008,Latif2016,Chon_and_Omukai2020}, and either intense Lyman–Werner (LW; $11.2 - 13.6~\mathrm{eV}$) radiation \citep[e.g.,][]{Omukai2001,Johnson_and_Bromm2007,Dijkstra2008,Shang2010,Johnson2011,Wolcott-Green2011,Agarwal2012,Safranek-Shrader2012,Latif2013c,Regan2014,Luo2020} or sufficiently high mass inflow rates to drive dynamical, compressional heating and fuel the accretion of the SMS phase \citep[e.g.,][]{Begelman2006,Begelman2008,Begelman2010, Hosokawa2012,Hosokawa2013,Latif2013e,Schleicher2013,Chon2018,Matsukoba2019}. Both LW radiation and inflow rates help to keep the gas close to the atomic-cooling temperature floor and thereby suppress the small-scale fragmentation that would otherwise be triggered by the efficient $\rm H_2$ cooling \citep[e.g.,][]{Yoshida2003,Wise2019,Lupi2021}.}

By modelling DCBH formation in cosmological hydrodynamical simulations, \citet{Cenci_and_Habouzit2025} show that the demographics of newly formed DCBHs, set by the physical conditions in their host nursery haloes, can naturally reproduce the observed redshift evolution in the emergence of LRDs \citep[see also][who reached similar conclusions]{Jeon2026}. In this picture, LRDs correspond to the first $10-100$ Myr after DCBH formation, during which a residual compact, dense, and quasi-pristine gaseous cocoon both enshrouds the accreting BH and suppresses the escaping X-ray emission. Recent work by \citet{Baggen2026} argues that more than about $40\percent$ of LRDs are consistent with having an UV-bright companion at an average projected physical separation of $\lesssim 5\kpc$. The fraction raises to $85\percent$ for the brightest LRDs in their sample. These findings further support the scenario where some LRDs may be associated with young DCBHs, where monolithic collapse is favoured by intense LW flux from the companion.

In this work, we focus on the possibility that a subset of high-$z$ LRDs are young systems hosting newly formed DCBHs. In this scenario, the red optical continuum is primarily associated with an accreting, enshrouded DCBH, while the UV emission is increasingly supplied by the first stellar populations assembling in the host galaxy. The observed UV diversity of LRDs may therefore encode an evolutionary sequence, in which the host contribution grows rapidly during the first few tens of Myr after DCBH formation. We use the \simname simulations to test whether the growth of newborn DCBHs and their host galaxies produces a predictable evolution of high-$z$ LRDs.

\section{Methods}\label{sec:methods}
\NEW{The \simname suite of simulations implements and tests different models for the formation of DCBHs in a cosmological framework, based on local properties of their host haloes and their large-scale environments. In this section, we describe the relevant properties of the one simulation from the \simname suite that we use in the present work. We refer the reader to Appendix~\ref{app:sim} for the full simulation setup, and to the upcoming methods papers (Cenci et al., in preparation) for a more detailed description of the full simulation suite.}

\subsection{The Simulation}
\NEW{For this work, we make use of a preliminary run part of the \simname suite, consisting of a cosmological (comoving) volume of $\left(\boxsizeL\cMpc\right)^3$ with periodic boundary conditions, evolved to a provisional final redshift of $z=\finalz$. The simulation start from a minimum refinement level $\ell_{\rm ref, min}=\levelmin$ (coarse level), corresponding to a comoving coarse resolution of $\Delta x_{\rm max}\simeq\spatialresmax\kpc$ and a dark matter mass resolution of $m_{\rm DM}\simeq 2\times 10^5 \Msun$. The stellar mass resolution of the present simulation is $m_{\star,0}\simeq \mstarres\Msun$. We adaptively refine the grid to a maximum refinement level $\ell_{\rm ref, max}=\levelmax$, achieving a maximum \NEW{proper} spatial resolution (i.e., the physical cell size at $\ell_{\rm ref, max}$) of $\Delta x_{\rm min}\simeq\spatialresmin\pc$.}

\NEW{A halo is eligible for DCBH formation in \simname if it meets \textit{all} of the following criteria around the halo's peak:
\begin{itemize}
    \item[($i$)]
    The gas density exceeds the threshold density for star-formation ($n>n_{\rm SF}$).
    \item[($ii$)]
    The gas-phase metallicity is below a critical value ($Z<10^{-4}~Z_\odot$).
    \item[($iii$)]
    The received Lyman-Werner flux exceeds the critical threshold $J_{\,\mathrm{LW}}^{\,\rm crit}$ (Equation~\ref{eqn:J_LW_crit}).
    \item[($iv$)]
    The mass inflow rate, $\dot{M}_{\rm in}$, exceeds the critical threshold $\dot{M}^{\,\rm crit}_{\,\rm in}$ (Equation~\ref{eqn:Mdot_crit}).
    \item[($v$)]
    The halo must not have formed stars yet: $M_{\star} = 0$.
\end{itemize}}

\NEW{If a halo is eligible for DCBH formation, then a DCBH is spawned stochastically at the halo's density peak. Seed DCBHs in have their mass determined by their host halo local properties (see Equation~\ref{eqn:BH_ini} in Appendix~\ref{sec:DCBH_model_methods}), rather than using a fixed seed mass value. The bolometric luminosity $L_{\rm bol}$ associated with the accretion process onto a DCBH is estimated from its accretion rate according to Equation~\eqref{eqn:Lbol_accr}, with a radiative efficiency set to $\effrad=0.1$.}

\subsection{SED modelling}\label{sec:SED_methods}
We consider a composite spectral energy distribution (SED) consisting of contributions from an accreting DCBH and PopIII stars. The DCBH emission is computed assuming a multi-blackbody spectrum from an optically thick, geometrically thin, $\alpha$-accretion disc \citep{Shakura_and_Sunyaev1973} surrounding a Schwarzschild black hole, given its mass and instantaneous mass-flow rate throughout the disc from the simulation output. The latter is estimated via the accretion rate in Equation~\eqref{eqn:Bondi}. The host-galaxy contribution is modelled using zero-age PopIII stars spectra \citep[from][]{Schaerer2002}, for varying DCBH-to-stellar mass ratios from the simulation. We account for effective UV attenuation from Rayleigh scattering with an optical depth \citep[see, e.g.,][]{Lee2013}:
\begin{equation}
    \tau_{\rm scatt} = \left(\frac{1216~\AAA}{\lambda_{\rm rest}}\right)^{4}\,\left(\frac{N_{\rm HI}}{4\times 10^{23}\,\mathrm{cm}^{-2}}\right)
    ~.
\end{equation}
\noindent Furthermore, we consider the following dust extinction law \citep[e.g.,][]{Temple2021}:
\begin{equation}
    \tau_{\rm dust} = \frac{A_{\rm V}}{1.086\,R_{\rm V}}\left[R_{\rm V}\,+\,6.2\,+\,\left(\frac{800~\AAA}{\lambda}\right)\right]
    ~,
\end{equation}
\noindent with an extinction ratio $R_{\rm V}=2.7$ \citep[typical value derived for the Small Magellanic Cloud and low-metallicity systems; e.g.,][]{Prevot1984,Calzetti2000}, and visual extinction given by the local neutral hydrogen column density $N_{\rm HI}$ and gas-phase metallicity $Z$ \citep[e.g.,][]{Bohlin1978}:
\begin{equation}
    A_{\rm V} = \left(Z/Z_\odot\right)\,\left(N_{\rm HI}/1.8\times 10^{21}~\mathrm{cm}^{-2}\right)
    ~.
\end{equation}
\noindent Dust extinction only plays a minor role in the overall attenuation, compared to Rayleigh scattering, for $Z\lesssim 10^{-6}~Z_\odot$. We assume that the accreting DCBH is enshrouded in a dense reservoir with column density $N_{\rm HI}\simeq 5\times 10^{25}~\mathrm{cm}^{-2}$ \citep[similar to, e.g.,][]{Pacucci2026} and pristine gas-phase metallicity $Z\simeq 5\times 10^{-5}~Z_\odot$, whereas for PopIII stars we assume they reside in a more diffuse environment with $N_{\rm HI}\simeq 10^{20}~\mathrm{cm}^{-2}$ and metallicity given by the simulation output (typically, $Z\gtrsim 1\times 10^{-5}~Z_\odot$). \NEW{We note that our results are mostly insensitive to the assumed value for $N_{\rm HI}\lesssim 10^{22}\,\mathrm{cm}^{-2}$ around PopIII stars.}

In this work, we primarily focus on an estimate for the UV-optical colour of our sources given by the emergent luminosity ratio $\deltamag= \log_{10}\left(L_{5100}/L_{2500}\right)$, between the emission\footnote{With $L_{5100}$ and $L_{2500}$ we indicate the luminosity $\lambda\,L_{\lambda}$ evaluated at a wavelength of $\lambda=5100~\AAA$ and $\lambda=2500~\AAA$, respectively.} at $0.51~\mu\mathrm{m}$ and $0.25~\mu\mathrm{m}$. In order to show data for non-LRD candidates in the simulation, where the emission of the DCBH accretion disc is drowned in that of PopIII stars, we separately compute $L_{5100}$ and $L_{2500}$ only accounting for the luminosity of the accretion disc and PopIII stars, respectively. This results in simulated sources with $\deltamag<0$. Although the inclusion of bound-free ($n=2$) Balmer continuum absorption introduces a more abrupt spectral break near $\lambda_{\rm rest}\simeq 0.37~\mu\mathrm{m}$, it does not affect our results or conclusions.

\NEW{The SED model above treats the entire host stellar population as zero-age PopIII, even though PopIII SNe (occurring from $3\Myr$ after their formation in the simulation) can enrich gas above our PopII threshold ($Z/Z_\odot>10^{-4}$). These metallicities are indeed measured in our simulated haloes, well within the $\lesssim30\Myr$ window that defines the early and late LRD phases (see below). We argue here that the resulting error in the predicted $0.2$-$0.6\,\mu{\rm m}$ colour is small. The spectral signatures that most clearly distinguish metal-free from higher-metallicity young stellar populations lie blueward of the $0.2$-$0.6\,\mu\mathrm{m}$ band considered here \citep[see][]{Schaerer2002}.}

We focus on the evolution of $\deltamag$ in simulated haloes hosting DCBHs in relation to DCBH ages, defined as the time since DCBH formation in the simulation. While the absolute values of $\deltamag$ depend on the assumed metallicity and HI column density around the DCBH, the qualitative evolutionary trends remain robust to variations in these parameters, provided that $N_{\rm HI}\gtrsim 10^{25}~\mathrm{cm}^{-2}$. Our conclusions are therefore based on the relative evolution of UV-optical colour, whereas a quantitative comparison with observed LRDs will require dedicated calibration.

\begin{figure*}
    \centering
    \includegraphics[width=\hsize]{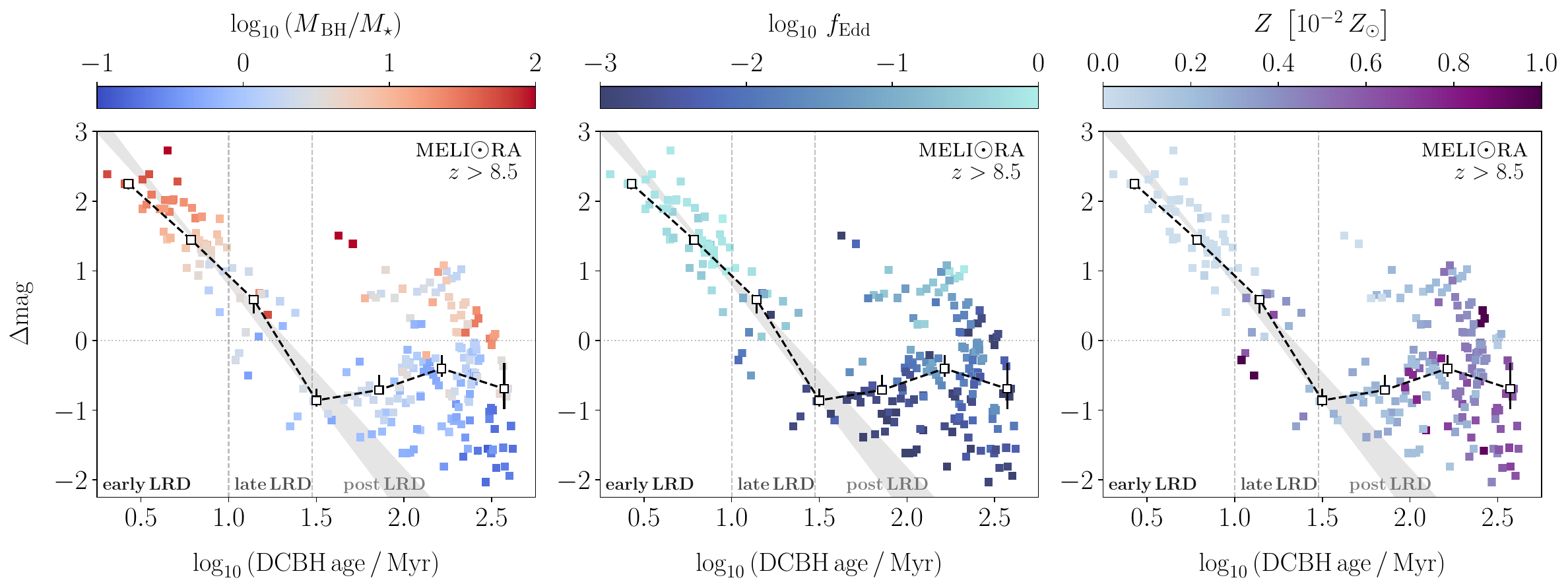}
    \caption{Evolution of the UV–optical colour $\deltamag=\log_{10}\left(L_{5100}/L_{2500}\right)$ for simulated DCBH+host systems at $z>8.5$ in \simname, as a function of the DCBH age (time since formation). Here, we only consider accreting DCBHs with bolometric luminosity $L_{\rm bol}>10^{40}~\mathrm{erg\,s^{-1}}$. The black dashed lines trace the median trend of $\deltamag$ as a function of DCBH age, with bootstrapped (16th to 84th percentiles) uncertainties. From left to right panels, we colour-code the data points by the BH-to-stellar mass ratio ($M_{\bh}/M_\star$), Eddington ratio ($f_{\rm Edd}$), and gas-phase metallicity ($Z$). \NEW{The shaded band encloses the 16th–84th percentile range for the best-fit linear relation, from bootstrap resampling of individual DCBHs, derived for $\mathrm{DCBH~age}<30\Myr$ $\deltamag>0$.} Immediately after the DCBH forms, the UV continuum is weak (high $\deltamag$) because stellar mass is negligible and the accretion flow onto the DCBH is just igniting. As the system ages, sustained star formation rapidly boosts the host’s UV luminosity and promotes metal enrichment, driving a steady decline in $\deltamag$ over the first $\sim 30\Myr$. The rapid evolution of $\deltamag$ implies that LRDs associated with newborn DCBHs are necessarily short-lived. In general, the relative UV-to-optical emission provides an age sequence for LRD-candidate DCBHs in their first $\sim 30\Myr$ after their formation.}
    \label{fig:delta_mag_vs_age}
\end{figure*}

\begin{figure*}
    \centering
    \includegraphics[width=\hsize]{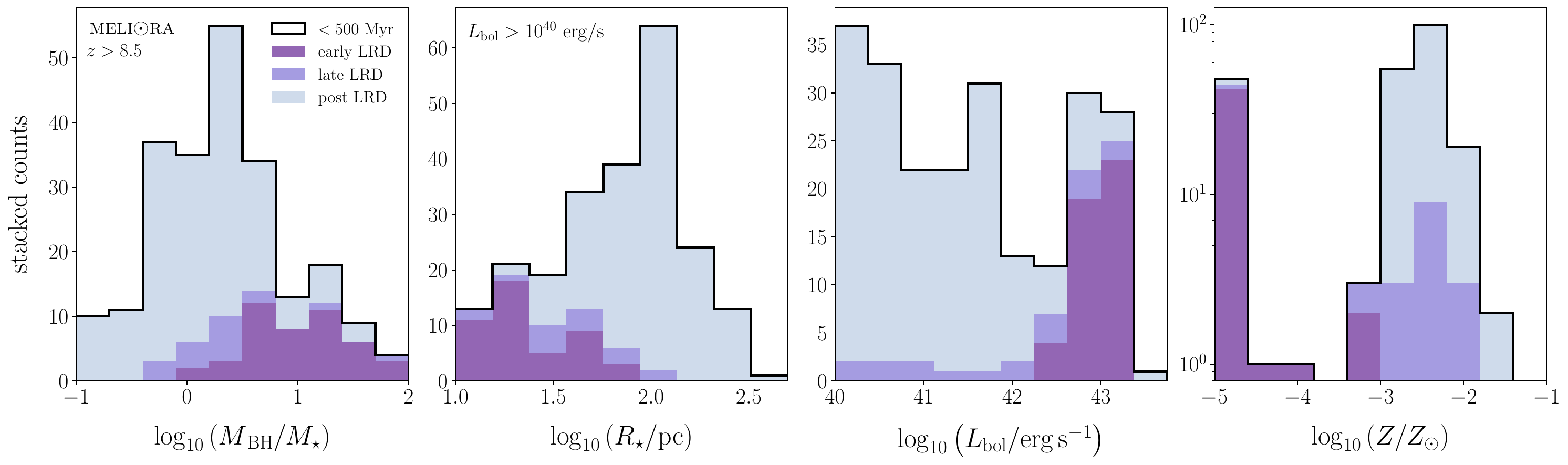}
    \caption{Stacked histograms for (from left to right) the BH-to-stellar mass ratio ($M_{\bh}/M_\star$), total extent of the stellar mass distribution ($R_\star$), bolometric luminosity ($L_{\rm bol}$), and gas-phase metallicity ($Z$), associated with DCBHs in the early, late, and post LRD phases. Each histogram bin stacks entries from all the different sub-categories considered. The solid black line shows the histogram of all DCBHs with ages $<500\Myr$. The same DCBH can contribute to different bins or sub-classes depending on the time at which it is analysed (effectively, depending on its age). Here, we only consider accreting DCBHs with $L_{\rm bol}>10^{40}~\mathrm{erg\,s^{-1}}$. In general, newborn DCBHs in early and late LRDs at $z>8.5$ in \simname naturally exhibit more compact stellar components, higher bolometric luminosities, and larger BH-to-stellar mass ratios than in post LRDs. As the DCBH ages and stars form within their host galaxy, the BH-to-stellar mass ratio rapidly decreases. Early and late LRDs have sizes below $\sim 100\pc$, before expanding significantly in the $>30\Myr$ after DCBH formation. Bolometric luminosities of early and late LRDs are often as high as $\sim 10^{43}~\mathrm{erg\,s^{-1}}$, to then decrease with DCBH age.}
    \label{fig:histos_age_bins}
\end{figure*}

\section{Results}\label{sec:results}
\NEW{We analyse 60 simulation outputs spanning $z\simeq 18$ to $z\simeq \finalz$, linearly spaced in expansion factor $a_{\rm exp}$, with an average separation of $\Delta a_{\rm exp}=0.001$ between consecutive outputs. Over this interval of cosmic time, a total of 65 DCBHs form within the simulated volume of $\left(\boxsizeL\cMpc\right)^3$. In the present analysis, however, we consider only those DCBHs that have had sufficient time to develop a stellar component ($M_\star>0$), allowing us to compare the relative contributions of the DCBH and stellar emission to the composite emergent SED. The number of DCBHs with a non-negligible stellar mass in their host galaxy increases with decreasing redshift, up to 35 DCBHs by $z\simeq 8.5$. In the simulation analysed in this work, the first DCBH forms at $z\simeq 13.7$, and the first DCBH with stellar host emerges at $z\simeq 13.5$.}

We estimate the relative contribution of PopIII stars and the accreting DCBH in LRD candidates in \simname (at $z>8.5$), via the colour proxy $\deltamag\equiv \log_{10}\left(L_{5100}/L_{2500}\right)$ between the rest-frame emission at $0.51~\mu\mathrm{m}$ and $0.25~\mu\mathrm{m}$. Figure~\ref{fig:delta_mag_vs_age} shows the evolution of $\deltamag$ as a function of the age of DCBHs in the simulation.

Immediately after DCBH formation, the absence of a significant stellar population results in weak UV emission, while accretion onto the DCBH is already in place, yielding large $\deltamag$ values. As the system evolves over the first few Myr, sustained star formation rapidly enhances the UV luminosity, while metal enrichment increases dust opacity and suppresses the direct UV contribution from the AGN, leading to a monotonic decrease in $\deltamag$ with BH age. The transition from BH-dominated to host-dominated emission is accompanied by a growing dispersion in $\deltamag$, possibly driven by variations in star formation efficiency, metal production, and accretion histories. Within this framework, we identify three main phases that LRD candidates may undergo:
\begin{itemize}

    \item[($i$)]
    Early LRD phase ($\mathrm{DCBH~age}<10\,\mathrm{Myr}$; with median $\deltamag\gtrsim 1$ in our model) -- immediately after DCBH formation; characterised by negligible stellar mass and weak UV emission from the host, pristine low gas-phase metallicities, and high Eddington ratios (possibly super-Eddington) for the accreting DCBH.
    
    \item[($ii$)]
    Late LRD phase ($10\,\mathrm{Myr}< \mathrm{DCBH~age}< 30\,\mathrm{Myr}$; with median $1\gtrsim \deltamag\gtrsim 0$ in our model) -- sustained star formation rapidly increases the UV luminosity of the host, while progressive metal enrichment enhances dust opacity. This phase is characterized by declining Eddington ratios, decreasing BH-to-stellar mass ratios of order unity, and rising gas-phase metallicities.

    \item[($iii$)]
    Post LRD phase ($\mathrm{DCBH~age}> 30\,\mathrm{Myr}$; with median $\deltamag\lesssim 0$ in our model) -- the stellar population dominates the UV output, while the accreting DCBH contributes primarily at longer wavelengths. In this regime, $\deltamag$ can reach much lower values and the system eventually no longer satisfies the conditions required for LRD selection. These objects represent a more evolved stage of early DCBH growth.
    
\end{itemize}

In Figure~\ref{fig:histos_age_bins}, we show the stacked histograms of the BH-to-stellar mass ratios, sizes of the stellar component, bolometric luminosities, and gas-phase metallicities, for DCBHs at $z>8.5$ in \simname, for the three different LRD phases defined above. Overall, the increasing contribution of the host galaxy, as measured by the decreasing BH-to-stellar mass ratio from early to post LRDs in \simname, is associated with: ($i$) higher metallicities, primarily due to enrichment from PopIII stellar evolution; ($ii$) reduced bolometric luminosities for the accreting black hole, as a result of BH feedback; ($iii$) increased sizes of the stellar component, that become larger than the characteristic $\sim 100\pc$ size of LRDs.

We predict that LRDs with strong UV emission trace a more evolved phase of early BH growth, while UV-faint systems correspond to younger BHs shortly after formation. These results suggest that the relative UV contribution in LRDs provides a physically motivated observational proxy for DCBH age during the first few tens of Myr of evolution. The rapid evolution of $\deltamag$ over $\lesssim 30\Myr$ implies that the LRD phase is intrinsically short-lived \citep[consistent with the findings of][]{Perez-Gonzalez2026}, placing strong constraints on both the emergence rate of LRDs and their redshift distribution.

We perform a linear regression \NEW{(see shaded area in Figure~\ref{fig:delta_mag_vs_age})} over the combined early+late LRD phases, restricting the sample to DCBHs with ages $t_{\bh}\leq 30\Myr$ and $\deltamag\geq 0$, and obtain:
\begin{equation}
    \log_{10}\,\frac{t_\mathrm{BH}}{\mathrm{Myr}} \,=\, -0.34^{+0.03}_{-0.04}\,\log_{10}\frac{L_{5100}}{L_{2500}} \,+\, 1.30^{+0.06}_{-0.04}
    ~.\label{eqn:age_vs_delta_mag}
\end{equation}
\noindent Similarly, we find:
\begin{equation}
    \log_{10}\,\frac{M_\mathrm{BH}}{M_\star\,L_{\rm bol,43}} \,=\, 1.19^{+0.03}_{-0.03}\,\log_{10}\frac{L_{5100}}{L_{2500}} \,-\,1.75^{+0.08}_{-0.07}
    ~,\label{eqn:massratio_vs_delta_mag}
\end{equation}
\noindent where $L_{\rm bol}\equiv L_{\rm bol,43}\,10^{43}~\mathrm{erg\,s^{-1}}$. The quoted best-fitting (median) parameters and uncertainties (16th–84th percentiles) are obtained via bootstrap resampling over individual DCBH histories, such that each bootstrap realisation removes or resamples entire BH evolutionary tracks rather than individual snapshots. The combined BH+stellar luminosities are computed assuming a dust attenuation and scattering model; however, our results are sensitive to the specific choice of dust law and column density. The most robust use of the relations above is likely differential rather than absolute: LRDs with larger $L_{5100}/L_{2500}$ should preferentially correspond to younger and more BH-dominated systems, while the more UV-bright sources should trace later stages in which the stellar component has grown.

\begin{figure}
    \centering
    \includegraphics[width=\hsize]{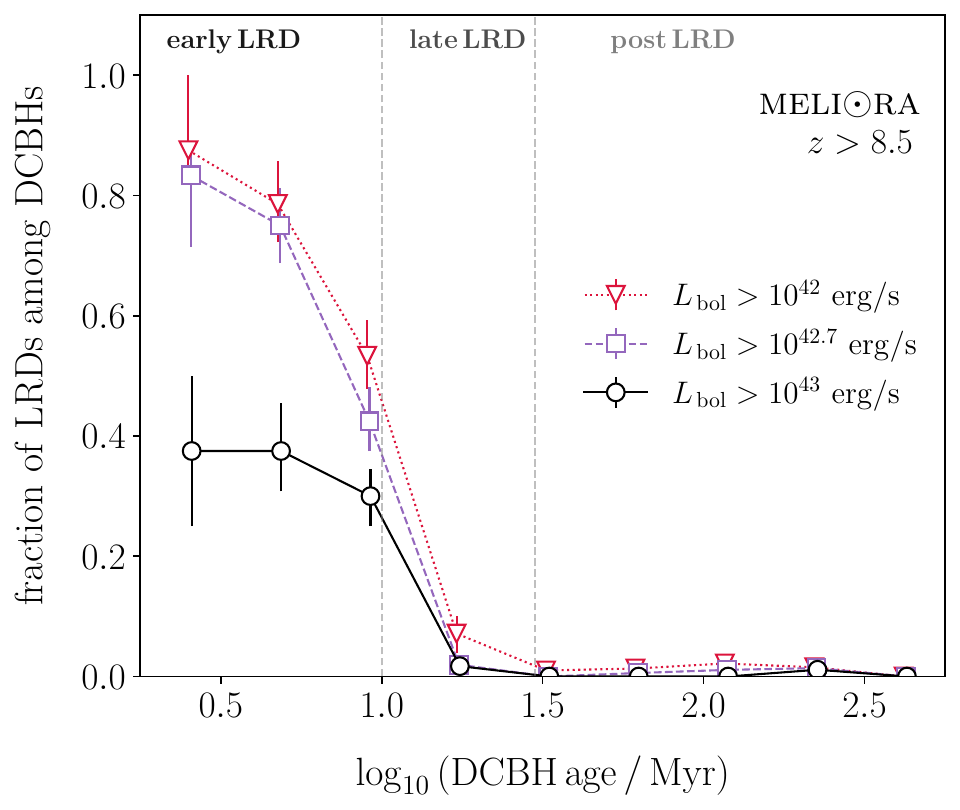}
    \caption{Fraction of DCBHs of a given age that are selected as LRD candidates in \simname, i.e. that have ($i$) compact stellar sizes $R_{\star}<100\pc$, ($ii$) large UV-optical colours with $\deltamag>0$, and ($iii$) bolometric luminosity above three different thresholds $L_{\rm bol}>10^{42},10^{42.7},10^{43}~\mathrm{erg\,s^{-1}}$ (dotted, dashed, and solid lines, respectively). Data points mark the median LRD fraction, with bootstrapped (16th to 84th percentiles) uncertainties. In general, the fraction of LRD candidates selected among newborn DCBHs decreases with DCBH age, consistently reaching zero for ages corresponding to the post-LRD phase.}
    \label{fig:LRD_frac_select}
\end{figure}

The abundance of newborn DCBHs (DCBH ages $\lesssim 100\Myr$) in \simname, at $8\lesssim z\lesssim 15$, ranges from $10^{-3}$ to $10^{-1}~\mathrm{cMpc}^{-3}$, depending on the specific DCBH formation model employed in the simulation \citep[see also][]{Cenci_and_Habouzit2025}. For the simulations used in the present work, the abundance of newborn DCBHs with $L_{\rm bol}>10^{40}~\mathrm{erg\,s^{-1}}$ is about $10^{-3}~\mathrm{cMpc}^{-3}$ at all redshifts $z\lesssim 15$. In Figure~\ref{fig:LRD_frac_select} shows the fraction of DCBHs of a given age that are selected as LRD candidates in \simname, i.e. that have ($i$) compact stellar sizes $R_{\star}<100\pc$, ($ii$) large UV-optical colours with $\deltamag>0$, and ($iii$) bolometric luminosity above three given thresholds $L_{\rm bol}>10^{42},10^{42.7},10^{43}~\mathrm{erg\,s^{-1}}$. In general, the fraction of LRD candidates selected among newborn DCBHs steeply decreases with DCBH age, from about $90\percent$ ($40\percent$) for early LRDs, to $\lesssim 20\percent$ ($\lesssim 10\percent$) for late LRDs, with a bolometric luminosity threshold of $10^{42}~\mathrm{erg\,s^{-1}}$ ($10^{43}~\mathrm{erg\,s^{-1}}$). Using a threshold $L_{\rm bol}>10^{40}~\mathrm{erg\,s^{-1}}$ yields an LRD fraction that is qualitatively the same as that with $L_{\rm bol}>10^{42}~\mathrm{erg\,s^{-1}}$. As a result, DCBH with ages $\lesssim 30\Myr$, with a duty cycle of about $30\percent$ for newborn DCBHs, we expect an abundance of about $10^{-4}~\mathrm{cMpc}^{-3}$, in line with some reported abundances for LRDs \citep[e.g.,][]{Kocevski2025}.

\NEW{Finally, we note that a small subset of post-LRD systems retains $\deltamag >0$ and $M_{\bh}>M_\star$, due to a relatively slow growth of their stellar host in mass, and could in principle be confused with early-phase LRDs on UV-optical colour alone. As shown in Figure~\ref{fig:histos_age_bins}, these systems have already developed extended stellar components (with $R_\star>100\pc$) by this stage, and are therefore excluded by the compactness criterion used in our LRD selection (see Figure~\ref{fig:LRD_frac_select}). The source size diagnostic thus breaks the degeneracy of colour and mass-ratio alone.}

\section{Discussion and conclusions}\label{sec:conclusions}
In this work, we use the \simname cosmological simulations to explore whether young, accreting DCBHs and their host galaxies can reproduce the UV-optical diversity of high-$z$ LRD candidates. In the following, we summarize the main implications of this scenario and highlight its key caveats, observational predictions, and future tests.

\subsection{BH-to-stellar mass ratios}
Newborn DCBHs in our model are expected to have very high BH-to-stellar mass ratios, because, by construction, the seed forms before the host has assembled a substantial stellar component. This makes our scenario most applicable to the youngest and most BH-dominated subset of LRDs, rather than to the full observed population. Systems with large inferred stellar masses or low $M_{\bh}/M_\star$ ratios may instead correspond to, e.g., later evolutionary stages or different seeding channels.

The inferred BH-to-stellar mass ratios of LRDs remain debated, but some recent studies point toward at least a sub-population of LRDs with overmassive BHs relative to their hosts \citep[e.g.,][]{Jones2026,Juodzbalis2025,Li2025,Matthee2025}. These measurements are particularly relevant for heavy-seed models, although they remain sensitive to virial mass calibrations, AGN-host decompositions, and assumptions about the origin of the broad-line widths. Recent stacked-SED analyses \citep[e.g.,][]{Perez-Gonzalez2026} suggest a range of host/AGN ratios, indicating significant diversity and selection biases in LRD samples. Heavy-seed (such as DCBH) models predict extreme early BH dominance, but their brief lifetimes and obscuring cocoons make them rare and hard to identify.

\subsection{SED modelling}
\citet{Perez-Gonzalez2026} interpret LRDs as compact galaxies hosting an intermediate–mass BH (within the theoretical DCBH mass range) together with an intense starburst. Their SED modelling combines stellar population templates with an AGN component (accretion disc plus dusty torus). They infer BH-to-stellar mass ratios of $\sim 1\percent$ and report strong broad FeII emission and Wolf–Rayet features, indicative of very young stellar populations embedded in dense gas. By contrast, \citet{Pacucci2026} model LRDs as isolated DCBHs without any accompanying galaxy,  simulate super-critical accretion onto a $10^5\Msun$ BH forming in a pristine halo, compute the emergent spectrum via full radiative transfer. The resulting spectrum exhibits a characteristic V-shaped SED with a pronounced Balmer break. Crucially, this model reproduces observed LRD spectra without invoking any stellar contribution, with intrinsically mild dust attenuation and negligible X-ray emission. In this framework, the LRD turnover arises purely from reprocessed AGN light \citep[see also, e.g.,][]{Greene2024,Labbe2024}, with the Balmer break dominated by bound-free absorption and no prominent hot-dust bump. Current observations are insufficient to discriminate between these two broad scenarios. While the models of \citet{Perez-Gonzalez2026} and \citet{Pacucci2026} represent rather opposite interpretations for the physical origin of LRDs, they are affected by substantial uncertainties in dust content, the stellar initial-mass function, and the BH accretion state. 

Future targeted observations of far-infrared/sub-mm bands, diagnostic nebular-line ratios, and deep X-ray stacking will be essential to break this degeneracy. Our model lies between these two extremes, as we allow for the presence of a stellar component in addition to the prominent contribution from the DCBH. Within our framework, these two interpretations may correspond to different stages of the same early evolutionary sequence. At the earliest times, the DCBH is embedded in a dense, nearly pristine cocoon and the stellar component is negligible, making the system closer to a DCBH-only SED. The host then rapidly contributes to the UV light and metal enrichment, and the system moves toward a mixed AGN+stellar phase. Eventually, the stellar component dominates the UV continuum and the source no longer satisfies the LRDs' colour selection. The SED diversity of LRD may therefore reflect both intrinsic differences between individual objects and rapid time evolution during the first $\sim 30\Myr$ after DCBH formation.

Our SED modelling is deliberately simplified. We do not perform full radiative transfer through a multi-phase, anisotropic medium, nor do we model the detailed geometry of the obscuring gas, nebular emission, resonant absorption, or dust formation. Consequently, the absolute values of $L_{5100}/L_{2500}$ should not be interpreted as precise predictions for observed LRD colours. A quantitative calibration of the proposed DCBH evolutionary sequence will require dedicated radiative transfer and forward-modelling through JWST filters, and we defer this task to future work.

\subsection{Interpretation}
We propose that, for at least a subset of LRDs, the relative contribution of the host galaxy and the enshrouded accreting BH evolves rapidly during the first \NEW{$\sim 30\Myr$} after DCBH formation, producing an age sequence encoded in $L_{5100}/L_{2500}$. Furthermore, this evolution is systematically associated with trends in emission line properties \citep[as reported by, e.g.,][]{de_Graaff2025,Matthee2026}, gas-phase metallicities, and BH accretion state.

The earliest LRD phase that we define is intrinsically rare because it lasts only a few Myr. However, the probability that a DCBH is selected as an LRD may be highest during this short interval, when the DCBH is still embedded in the compact gas reservoir associated with the DCBH formation event and the stellar UV contribution remains weak. Later stages can last longer, but the growing host galaxy progressively dilutes the red AGN continuum and lowers the probability of satisfying strict LRD colour and morphology cuts. While older systems naturally move toward mixed AGN+stellar interpretations, earlier LRDs may resemble the DCBH-only spectra proposed by \citet{Pacucci2026}, with prominent Balmer breaks and weak nebular emission \citep[as observed in sources like, e.g.,][]{Naidu2025}. Finally, some of the compact, UV-bright descendants of early LRD systems (in their late or post LRD phases) could be naturally associated with the population of so-called Little Blue Dots \citep[LBDs;][]{Brazzini2026}.

Merger-driven BH formation channels may produce heavy seeds in environments that are already enriched and dynamically complex \citep[e.g.,][]{Mayer_and_Bonoli2019,Mayer2024}. Such systems may still follow a qualitatively similar decline in the AGN contribution as the host grows, but they would not necessarily exhibit the same pristine early phase or metallicity evolution predicted by our simulation.

An important alternative, or complementary, interpretation is that part of the observed LRD diversity is driven by projection effects rather than BH age. If the obscuring medium is clumpy, flattened, or otherwise non-spherically symmetric, different viewing angles can produce different UV-optical colours at fixed intrinsic BH age. This picture is qualitatively consistent with recent interpretations in which LRD spectral diversity reflects variations in the geometry, covering factor, and column density of dense gas around the accreting BH \citep[e.g.,][]{Brazzini2026,Matthee2026}.

\subsection{Limitations}
 \NEW{As noted above, the fact that DCBHs are extremely overmassive immediately after formation are imposed by the model rather than being an independent prediction of it. A genuine prediction of the simulation is the subsequent rate at which this ratio declines as the host assembles stars, which follows self-consistently from the star-formation and feedback physics. Star formation is also regulated by black hole feedback, that is in turn affected by the initial choice DCBH mass, derived as function of local halo properties. Testing our predictions when relaxing the $M_\star = 0$ criterion is deferred to future work.}

\NEW{Our simulation does not resolve $\rm H_2$ cooling or star formation in haloes below the mass resolved by 50 dark matter particles (i.e., with halo masses $\lesssim 10^7\Msun$). PopIII-driven supernova remnants in such (mini-)haloes are known to enrich gas out to about a proper kpc scale, which can establish a metallicity floor of about $Z\sim 10^{-3}~Z_\odot$ in the smallest galaxies \citep[e.g.,][]{Wise2008b,Jeon2014,Ritter2012,Chen2024}. The mixing of this enriched gas into larger volumes could reduce the pristine gas fraction available for subsequent DCBH formation.}

\NEW{We note that we do not model radiative feedback from the intermediate supermassive-star (SMS) phase that precedes DCBH formation. Albeit effective only for their short ($\sim 1\Myr$) lifetime, SMSs are a strong source of IR radiation that could further suppress local $\rm H_2$ formation and possibly PopIII star formation in their immediate vicinity.}

\subsection{Future directions}
A key future test of our evolutionary picture will be to connect the UV-optical colours of high-$z$ LRDs to their resolved morphology, AGN-host decomposition, and gas-phase metallicity. In our scenario, UV-brighter and bluer systems should correspond to later stages, in which the stellar component has grown, metal enrichment has progressed, and the host galaxy begins to extend beyond the compact DCBH-forming region. In general, deep JWST imaging could test whether rest frame UV sizes, colour gradients, or the incidence of nearby UV-bright companions correlate with $L_{5100}/L_{2500}$. Similarly, rest frame optical emission line diagnostics could reveal whether bluer LRDs preferentially show stronger metal lines and higher inferred metallicities, as expected if they trace more evolved host galaxies. Strongly lensed systems are particularly valuable, possibly enabling deep spectroscopy and spatially resolved constraints on the host and AGN components. For instance, recent studies of lensed or magnified LRDs were able to constrain the low gas-phase metallicities of some systems \citep[e.g.,][]{Maiolino2026,Kokorev2025,Tripodi2025}.

Since the simulation used in this work is currently evolved only to $z\simeq 8.5$, our conclusions are mostly restricted to the earliest, high-$z$ LRD candidates will need to be extended to lower redshifts. In addition, the currently simulated volume of $(10~\rm{cMpc})^3$ is relatively limited, which may prevent us from sampling the rarest environments and the most massive BHs expected at these epochs. This work should be interpreted as a first exploration of the early phases of DCBH growth in representative high-$z$ environments. Future work carried out as part of the \simname simulation project, including larger volumes and evolution to lower redshift, will be essential to assess whether this evolutionary picture can extend to more massive systems and to the broader LRD population at later cosmic times.

\section*{Acknowledgements}
MH and EC acknowledge support from the Swiss SNSF Starting Grant (grant no. 218032). The computations were performed at the University of Geneva, using \textsc{yggdrasil} HPC service, and at the Department of Astronomy of UNIGE using \textsc{bonsai} HPC service. All plots were created with the \textsc{matplotlib} library for visualization with Python \citep{Hunter2007}.

\bibliography{main}
\bibliographystyle{mnras}

\newpage\appendix
\section{Simulation prescriptions}\label{app:sim}
\NEW{This appendix gives the full set of physical prescriptions underlying the \simname simulation used in this work, expanding on the condensed summary given in Section~\ref{sec:methods}. With the exception of the simulation volume and resolution, and a few details in the black hole formation model, which differ from that of \citet{Cenci_and_Habouzit2025} and is described here in full, the components summarised below are unchanged from that paper. We refer the reader to the upcoming methods papers (Cenci et al., in preparation) for details on the full \simname simulation suite. The SED-modelling framework, entirely new to this paper, is presented in full in Section~\ref{sec:SED_methods} of the main text rather than here.}

\subsection{Initial conditions}
Initial conditions are generated at $z=100$, using the MUlti Scale Initial Conditions \citep[MUSIC;][]{Hahn_and_Abel2011} code. We assumed cosmological parameters $\Omega_{\rm m}=0.272$, $\Omega_\Lambda=1-\Omega_{\rm m} = 0.728$, $\Omega_{\rm b}=0.045$, $h=0.704$, $\sigma_8=0.81$, $n_{\rm s}=0.967$, that are consistent with results from the Wilkinson Microwave Anisotropy Probe \citep[WMAP-7; measurements taken over 7 years of observations;][]{Komatsu2011}. We employed a \citet{Eisenstein_and_Hu1998,Eisenstein_and_Hu1999} transfer function for a cold dark matter cosmology, that includes perturbations due to several baryon effects.

\subsection{Simulation code}
Gravity and hydrodynamics in \simname are solved with the adaptive mesh refinement \ramses code\footnote{The publicly available repository for \ramses can be found at: \url{https://bitbucket.org/rteyssie/ramses/src/master/}.} \citep[][]{Teyssier2002}. \ramses is fundamentally based on a fully threaded oct-tree structure \citep[e.g.,][]{Kravtsov1997,Khokhlov1998}.

Euler equations for gas dynamics on the grid are solved with an unsplit, second-order MUSCL-Hancock scheme \citep[][]{van_Leer1979} and the Harten-Lax-Van Leer-Contact \citep[HLLC;][]{Toro1994,Toro1999} approximate Riemann solver. The dynamics of collisionless particles (dark matter, stars, and BHs) is solved for using a particle-mesh (PM) method with cloud-in-cell (CIC) interpolation. Time-steps of finer levels in the mesh hierarchy are updated twice for every update step in the next coarser levels. For stability, time-steps are limited following the Courant–Friedrichs–Lewy condition, with a Courant number of 0.8.

\subsection{Grid refinement}
The \simname simulations start from a minimum, coarse refinement level $\ell_{\rm ref, min}$. We adaptively refine the grid to a maximum refinement level $\ell_{\rm ref, max}$. Refinement is allowed over the entirety of the simulated volume. The criterion for mesh refinement follows a quasi-Lagrangian approach: for a given cell, refinement is triggered if
\begin{equation}
    M_{\rm DM} + \frac{\Omega_{\rm m}}{\Omega_{\rm b}}\,M_{\rm b} \,\geq\, 8\,m_{\rm DM}
    ~,
\end{equation}
\noindent where $M_{\rm DM}$ and $M_{\rm b}$ are the total dark matter and baryon (gas + stars) mass in the cell, respectively. Additionally, a cell is refined if its size is larger than twice the characteristic Jeans length computed from the cell's temperature and density. Each refinement level consists of a characteristic cell size that is half of that of the previous coarser level. The cells containing sink particles (representing, e.g., BHs) are enforced to get refined to the maximum refinement level that is currently allowed. To keep an approximately constant physical scale-length for grid cells at the finest refinement level, we allow for an additional refinement level (above $\ell_{\rm ref, min}$) every time the scale factor doubles.

\subsection{Clump finder}
Haloes in our simulations are associated with gas density clumps, that are identified throughout the simulations with the built-in clump-finder routines of \ramses \citep[][]{Bleuler2014,Bleuler2015}. Gas cells with a mass density above the density threshold $\hat{\rho}_{\rm peak}=80\,\rho_{\rm crit,\,m}$ are assigned to the nearest density peak, following the steepest-ascent path. Here $\rho_{\rm crit,\,m}=1.88\times 10^{-29}\Omega_{\rm m}\,h^2\,\left(1+z\right)^3~\left[\mathrm{g}/\mathrm{cm^{3}}\right]$ is the critical matter density of the Universe at the given redshift. Saddle points between peaks, where density is larger than $\hat{\rho}_{\rm saddle}=200\,\rho_{\rm crit,\,m}$, are also identified in order to merge the non-relevant structures to their nearby clumps. When the peak-to-saddle density ratio is below a relevance threshold $\chi_{\rm clump}=3$, the peaks are merged together. Otherwise, both peaks survive and are considered as independent clump structures.

\subsection{Gas cooling and heating}\label{sec:gas_cooling}
The equation of state of gas in \simname is that of an ideal gas with an adiabatic index $\gamma=5/3$ (monoatomic gas). Gas cooling follows the model by \citet{Sutherland_and_Dopita1993}, accounting for cooling by H, He, and metals, down to $10^4~\mathrm{K}$. At temperatures $\lesssim 10^4~\mathrm{K}$, we include fine-structure, infrared line cooling through the analytical approximation of the cooling rates by \citet{Rosen_and_Bregman1995} \citep[based on the results by][]{Dalgarno_and_McCray1972}. We assume an initial average gas-phase metallicity $Z_{0}=0$. After redshift $z_{\rm reion}=8.5$, we consider gas heating from an uniform UV background from high-$z$ reionization sources, following the model by \citet{Haardt_and_Madau1996}. \NEW{However, since the simulation analysed here only evolved down to $z\simeq 8.5$, this background is never actually active in any of the outputs used in this work, and gas heating is due solely to the polytropic temperature floor and stellar/BH feedback described below.} In order to mimic the effect of gas heating in star-forming regions, and prevent excessive spurious fragmentation, we impose a polytropic temperature floor $T_{\rm floor} = T_{0}\,\left(n/n_{\rm SF}\right)^{\kappa-1}$ of the high-density interstellar medium, where $\kappa=1.6$ and $T_{0}=10^3~\mathrm{K}$. At high densities, above the density threshold for star formation, $n_{\rm SF}$, this prescription results in an artificially increased thermal pressure that prevents gas from cooling to lower temperatures \citep[e.g.,][]{Springel_and_Hernquist2003}.

\subsection{Star formation}
Star formation in \simname takes place in gas with hydrogen density above a fixed threshold $n_{\rm SF}=\nsf~\mathrm{cm}^{-3}$. Gas cell satisfying $n > n_{\rm SF}$ can form a star particle, with an expected, average star-formation rate density $\dot{\rho}_{\star}$ following the \citet{Schmidt1959} law:
\begin{equation}
    \dot{\rho}_{\star} = \epsilon_{\rm ff}\,\frac{\rho}{t_{\rm ff}}
    ~,
\end{equation}
\noindent where $\rho$ is the cell's gas mass density, $t_{\rm ff}=\sqrt{3\pi/32\,G\,\rho\;}$ is the local free-fall time, and $\epsilon_{\rm ff}$ is the \textit{local} star formation efficiency per free-fall time. While often taken as a fixed parameter independent of redshift, here $\epsilon_{\rm ff}$ is estimated based on the local properties of the interstellar medium, following the turbulence-regulated model by \citet{Federrath_and_Klessen2012} and \citet{Padoan_and_Nordlund2011} \citep[see also][]{Hennebelle_and_Chabrier2011}.

Over a time-step $\Delta t$, a \NEW{parent gas cell forms a star particle at its centre} that represents a population of a number $N_\star$ of stars, that follows Poisson statistics, and limited such that every gas cell can convert up to a maximum $90\percent$ of their available gas mass into stars. The minimum resolved stellar mass in the simulations used in this work is $m_{\star,0}\simeq \mstarres\Msun$. The newly formed star particles have a total mass $m_\star=N_\star\,m_{\star,0}$ and inherit momentum ($p_\star$) and metallicity ($Z_\star$) from their parent gas cell.

Star particles with metallicity $Z/Z_\odot<10^{-4}$ are classified as PopIII stars, whereas PopII stars have larger metallicities \citep[see, e.g.,][]{Maio2011}. Following, e.g., \citet{Kimm2017}, PopIII stars in \simname form with a \citet{Salpeter1955} initial-mass function (IMF), whereas other stars form with a \citet{Kroupa2001} IMF.

\subsection{Stellar feedback}
We collectively consider feedback from type II and type Ia supernovae (SNe), that deposit thermal energy, kinetic energy, and metals in the surrounding medium. Furthermore, we distinguish between feedback from PopIII and PopII stars. Star particles that are older than $10\Myr$ undergo SN explosions that eject a fixed fraction $\eta_{\rm SN}=0.2$ of their mass. This is numerically equivalent to assuming that $20\percent$ of the stars simultaneously go off as SNe in the coeval population represented by simulated star particles. The ejected mass is instantaneously deposited in the 27 neighbouring father cells, with a metal yield of $Y_{\rm SN}=0.1$. Therefore, when triggering SN events, a star particle ejects a mass $M_{\mathrm {ej}}=\eta_{\rm SN}\,m_{\star}$, with a metallicity $Z_{\mathrm{ej}}=Z_{\star}\,+\,Y_{\rm SN}(1-Z_{\star})$. Furthermore, SNe release a total energy of $E_{\rm SN}=10^{51}~\mathrm{erg}$ per $10\Msun$ of ejected mass. In our simulations, we do not resolve the radius of SNe remnants at the end of the Sedov–Taylor phase and the energy ($E_{\rm SN}$) that is released is instantaneously deposited as thermal energy \NEW{into the same 27 neighbouring father cells used
for the mass and metal deposition described above}. In \simname, PopIII star particles can go off as SNe after only $3\Myr$ \citep[approximately the average lifetime for a $40\Msun$ PopIII star; see][]{Schaerer2002,Schaerer2003}. We assume an ejecta mass fraction of $\eta_{\rm SN}=0.35$, with a total energy release of $10^{51}~\mathrm{erg}$ per $10\Msun$ of ejected mass and a metal yield of $Y_{\rm SN}=0.2$ \citep[see][and references therein]{Wise2012a,Kimm2017}.

Massive young stars likely reside in dense gas that can efficiently cool, quickly radiating away the thermal energy deposited by SN feedback \citep[e.g.,][]{Katz1992}. However, non-thermal processes (e.g., unresolved turbulence, magnetic fields, and cosmic rays) can store additional energy that dissipates over time-scales that are typically longer than the gas cooling time-scale. In order to mimic the effect of non-thermal pressure terms, we turn off the cooling for gas around star particles for a dissipation time-scale $t_{\rm diss}=20\Myr$ \citep[see, e.g.,][]{Gerritsen1997,Thacker_and_Couchman2000,Thacker_and_Couchman2001,Stinson2006,Teyssier2013}, of the order of the lifetime of massive ($>10\Msun$) stars. In \simname, we use the same dissipation time-scale for both PopII and PopIII star particles.

\subsection{Lyman-Werner radiation}
The total (unattenuated) Lyman-Werner (LW) flux $J_{\,\mathrm{LW},i}$ at the centre of the $i$-th halo is computed accounting for an uniform LW background flux \citep[$J_{\,\rm LW}^{\,\rm bk}$; following][]{Incatasciato2023} and the LW flux received from star forming regions around the target halo:
\begin{equation}
    J_{\,\mathrm{LW},i} 
    \,=\, J_{\,\rm LW}^{\,\rm bk} \,+\,
    \sum_{n}^{\rm stars}\,\mathcal{A}_{\star, n}\left(\frac{m_{\star,n}}{10^3\Msun}\right)\left(\frac{\left\lvert\bm{r}_i-\bm{r}_{\star,n}\right\rvert}{1\kpc}\right)^{-2}
    ~.\label{eqn:J_LW_0}
\end{equation}
\noindent Here $m_{\star,n}$ and $\bm{r}_{\star,n}$ are the mass and position vector of the $n$-th star particle, respectively. The factor $\mathcal{A}_{\star,n}$ includes the dimming of the LW flux with increasing stellar age $t_\star$ \citep[][]{Schaerer2003,Lupi2021}. For the $n$-th star particle, we have:
\begin{equation}
    \mathcal{A}_{\star,n} = \frac{f_{\star,n}\,\exp\left[\,-t_{\star,n}/\left(300\Myr\right)\,\right]}{\left[\,1+t_{\star,n}/\left(4\Myr\right)\,\right]^{3/2}}
    ~,
\end{equation}
\noindent where $f_{\star,n}=15$ for PopIII stars, and $f_{\star,n}=3$ otherwise \citep[e.g.,][]{Greif_and_Bromm2006,Agarwal2014}.

We correct the LW fluxes of Equation~\eqref{eqn:J_LW_0} for attenuation and shielding,  \NEW{by estimating the local molecular gas fraction and column density of haloes along the line-of-sight of every source \citep[as in, e.g.,][]{Shang2010,Latif2015,Latif_and_Volonteri2015}, assuming that the H$^{-}$ channel is the dominant one for H$_2$ formation \citep[e.g.,][]{Yoshida2006b,Lupi2021}.}

\subsection{Mass-inflow rates}
We estimate the instantaneous mass-flow rate $\dot{M}_i$ onto the central region (where we expect DCBHs to form) of the $i$-th halo as follows:
\begin{equation}
    \dot{M}_{\mathrm{in},\,i} \,=\, -\sum_{n}\,\frac{m_{n}\,v_{r,n}}{r_n}
    ~,\label{eqn:Mdot}
\end{equation}
\noindent where $m_n$, $v_{r,n}$ and $r_n = \left\lvert\bm{r}_i-\bm{r}_{n}\right\rvert$ are the mass, radial component of velocity, and distance, respectively, of the gas in the $n$-th cell, with respect to the centre of the $i$-th halo.

\subsection{Black hole formation}\label{sec:DCBH_model_methods}
At each time-step, for all haloes identified in a given simulation, we also evaluate the physical conditions around the density peak, within a spherical accretion radius $r_{\rm accr} = 4\,\Delta x_{\rm min}$, where $\Delta x_{\rm min}$ is the size of a cell at the highest refinement level in the simulation. Note that the mass-inflow rate used in the criteria below is evaluated within this same radius $r_{\rm accr}$.

A given target halo is eligible for DCBH formation if all of the following criteria are met:
\begin{itemize}
    \item[($i$)]
    The average density, $\bar{n}$, around the halo's peak exceeds the threshold density for star-formation, $n_{\rm SF}=\nsf~\mathrm{cm^{-3}}$.
    
    \item[($ii$)]
    The average gas-phase metallicity, $\bar{Z}$, around the halo's peak must be below a fixed critical threshold, $Z^{\,\rm crit}=10^{-4}~Z_\odot$.
    
    \item[($iii$)]
    The effective Lyman-Werner flux, $J_{\,\rm LW}$, at the halo's peak must exceed a critical value, $J_{\,\rm LW}^{\,\rm crit}$ (see Equation~\ref{eqn:J_LW_crit}).

    \item[($iv$)]
    The mass-flow rate, $\dot{M}_{\rm in}$, toward the halo's density peak must exceed a critical value, $\dot{M}^{\,\rm crit}_{\rm in}$ (see Equation~\ref{eqn:Mdot_crit}).

    \item[($v$)]
    The halo must not have formed stars yet: $M_{\star} = 0$.
\end{itemize}

\NEW{We do not impose a lower halo-mass limit beyond the resolution floor implied by our dark matter particle mass ($m_{\rm DM}\simeq 2\times 10^{5}\Msun$; i.e. haloes are only meaningfully resolved above approximately $10^7\Msun$, corresponding to about 50 dark matter particles).}

Furthermore, we do not impose any upper limit on the spin of the dark matter halo. Halo spin it is not clearly coupled to the angular momentum of baryons on smaller scales \citep[e.g.,][]{Dubois2012b,Bonoli2014}. A low-spin criterion for DCBH formation is generally less restrictive than requiring, e.g., a minimum incoming LW flux \citep[see][]{Bhowmick2022a}. A number of combinations of criteria and thresholds are explored in the full \simname suite of simulations, as well as different resolutions and volumes.

\subsubsection{Local critical Lyman-Werner flux}
Assuming fully neutral (i.e., mean molecular weight $\mu=1.22$) monoatomic medium (adiabatic index $\gamma=5/3$) in a virialized halo, we can estimate the local, critical LW flux $J_{\,\mathrm{LW}}^{\,\rm crit}$ above which LW radiation prevents H$_2$ from efficiently forming \citep[see, e.g.,][and references therein]{Schauer2019,Kulkarni2021,Lupi2021}:
\begin{equation}
    \frac{J_{\,\mathrm{LW}}^{\,\rm crit}}{100\,J_{21}} \;=\; \left(\frac{1+z}{10}\right)^{1.8}\,\left(\frac{n}{\mathrm{cm}^{-3}}\right)^2\,\left(\frac{M}{10^{8}\,\mathrm{M}_\odot}\right)^{2.2}
    ~,\label{eqn:J_LW_crit}
\end{equation}
\noindent in units of $J_{21} = 10^{-21}$ erg s$^{-1}$ cm$^{-2}$ sr$^{-1}$ Hz$^{-1}$, where $n$ and $M$ are the average hydrogen density and total mass (dark matter + baryons) of the target halo at redshift $z$, respectively. The critical threshold $J_{\,\mathrm{LW}}^{\,\rm crit}$ is derived by asking that $\epsilon_{\rm H_2}\equiv t_{\rm cool, \rm H_2}^{\,\rm min}/t_{\rm H}=1$, i.e. that the minimum H$_2$ cooling time-scale $t_{\rm cool, \rm H_2}^{\,\rm min}$ allowed for DCBH formation is comparable to the Hubble time-scale $t_{\rm H}$. In the simulation used in the present work, we ask a target halo to satisfy $J_{\,\mathrm{LW}}>J_{\,\mathrm{LW}}^{\,\rm crit}$ in order to be eligible for DCBH formation.

\subsubsection{Dynamical heating}
Dynamical, compressional heating can counteract cooling and prevent H$_2$ formation for halo mass-growth rates above a local threshold value $\dot{M}^{\,\rm crit}_{\,\rm in}$ \citep[see, e.g.,][and references therein]{Yoshida2003,Wise2019,Lupi2021}. Furthermore, sustained inflow rates onto the central regions of DCBH-forming haloes are required to fuel the accretion of the intermediate protostar phase and overcome the action of radiative feedback \citep[e.g.,][]{Hosokawa2012,Hosokawa2013,Schleicher2013,Chon2018}. Following, e.g., \citet{Lupi2021}, for a target halo at redshift $z$ we have:
\begin{align}
   \frac{\dot{M}^{\,\rm crit}_{\,\rm in}}{\mathrm{M}_\odot/\mathrm{yr}}
   \,&=\,\left(\frac{1+z}{10}\right)^{3/2}\left(\frac{M}{10^8\,\mathrm{M}_\odot}\right)\left(\frac{J_{\,\mathrm{LW}}}{J_{\,\mathrm{LW}}^{\,\rm crit}}\right)^{-1}
   ~,\label{eqn:Mdot_crit}
\end{align}
\noindent where $n$, $M$, $J_{\,\mathrm{LW}}$ are the average hydrogen density, total mass (dark matter + baryons), and received LW flux of the target halo at redshift $z$, respectively. In order to be eligible for DCBH formation, the target halo must satisfy $\dot{M}_{\rm in}>\dot{M}^{\,\rm crit}_{\,\rm in}$.

\subsubsection{Seed mass}
The formation sites of DCBHs are expected to experience intense LW flux and mass inflow rates over a finite period of time of a few Myr. To account for this, as well as the accretion phase preceding the collapse of supermassive stars (i.e. the immediate progenitors of DCBH), DCBHs in \simname are spawned stochastically in haloes that are eligible for DCBH formation, according to the criteria listed above. At each coarse time-step $\Delta t$, we consider a Poisson probability $\mathcal{P}_{\bh}$ of forming a DCBH sink particle at the centre of the given halo over that time-step \citep[similarly to, e.g.,][]{Bellovary2011,Dunn2018}:
\begin{equation}
    \mathcal{P}_{\bh} = 1 - \exp\left(-\Delta t / t_{\mathrm{ff}}\right)
    ~,
\end{equation}
\noindent where $t_{\mathrm{ff}} = \sqrt{3\pi/32\,G\,\bar{\rho}\;}$ is the local free-fall time in the halo centre. For star-forming gas we have $t_{\rm ff}\lesssim 70\Myr$. Here $\bar{\rho}$ is the average gas mass density in the cells around the halo centre. We draw a random number $x\in \left[0,1\right]$ from an uniform distribution and, if $x<\mathcal{P}_{\bh}$, we spawn a sink particle representing a new DCBH forming in the simulation. Unlike most state-of-the-art simulations \citep[e.g., Illustris, EAGLE; but see][]{Habouzit2017}, the seed BHs in \simname have a mass $M_{\bh}^{\,\rm seed}$, that is determined by the local properties of their host haloes. Specifically, we compute $M_{\bh}^{\,\rm seed}$ as a function of the local inflow rate, following the predictions of recent theoretical models \citep[e.g.,][]{Umeda2016,Haemmerle2018,Woods2017,Woods2019}:
\begin{equation}
    \frac{M_{\bh}^{\,\rm seed}}{10^5\Msun} \,=\, 2.74\,+\,2.21\,\mathcal{X}\,+\,0.64\,\mathcal{X}^2
    ~,\label{eqn:BH_ini}
\end{equation}
\noindent where we defined:
\begin{equation}
    \mathcal{X}\,=\,\log_{10}\left[\frac{\langle\,\dot{M}_{\rm in}\,\rangle_{\bh}}{\mathrm{M}_\odot\,\mathrm{yr}^{-1}}\right]
    ~.
\end{equation}
\noindent Here $\langle\,\dot{M}_{\rm in}\,\rangle_{\bh}$ is the average mass  inflow rate in the vicinity of the BH formation site. Around the DCBH formation site, we also spawn a swarm of \textit{cloud} particles, \NEW{regularly} spaced by $\Delta x/2$ and distributed within a sphere of radius $r_{\rm cloud} = 4 \Delta x$, where $\Delta x$ is the local gas cell size. Cloud particles are implemented purely with a numerical purpose, to probe the average properties of gas surrounding BH particles. With $\left\langle\,\cdot\,\right\rangle_{\bh}$ we indicate a volume-weighted average computed using a cloud-in-cell (CIC) over the 8 nearest father cells around the sink particle, within $r_{\rm cloud}$. \NEW{We note that relation~\eqref{eqn:BH_ini} is only formally valid for high enough accretion rates $\langle\,\dot{M}_{\rm in}\,\rangle_{\bh}\gtrsim 10^{-1.727}~\mathrm{M}_\odot\,\mathrm{yr}^{-1}$. The accretion rate at the time of DCBH formation is generally above this lower limit, as it is enforced by the other formation criteria and the halo mass resolution of our simulations.}

\subsection{Black hole growth}
DCBHs in the \simname simulations grow due to both black hole mergers and gas accretion, which affect the evolution of their mass and spin.

The black hole mass growth rate due to gas accretion is estimated starting from the Bondi-Hoyle-Lyttleton \citep[][]{Hoyle_and_Lyttleton1939,Bondi_and_Hoyle1944,Bondi1952} solution:
\begin{equation}
    \dot{M}_{\rm B} \,=\,\frac{4\,\pi\,G^2\,M_{\bh}^2\,\left\langle\rho\right\rangle_{\bh}/\alpha_{\rm B}}{\left[\;\left\lvert \bm{v}_{\bh}-\left\langle\bm{v}\right\rangle_{\bh}\right\rvert^2 + \left\langle c^2_{\rm s}\right\rangle_{\bh}\;\right]^{3/2}}
    ~,\label{eqn:Bondi}
\end{equation}
\noindent where $M_{\bh}$, and $\bm{v}_{\bh}$ are the black hole mass and velocity, respectively, whereas $\bm{v}$, $c_{\rm s}$, and $\rho$ are the velocity, sound speed, and mass density of gas in the cells surrounding the sink particle, respectively. With $\left\langle \cdot\right\rangle_{\bh}$ we represent the volume-weighted average with CIC interpolation over the 8 nearest father cells around the sink particle, within $r_{\rm cloud}$. The correction factor $\alpha_{\rm B}$ is introduced to extrapolate the average gas quantities to their value far from the black hole \citep[see][]{Krumholz2004,Dubois2010}. 

Furthermore, we limit the accretion rate $\dot{M}_{\rm B}$ to the Eddington critical accretion rate $\dot{M}_{\rm Edd} = 4\pi\,G\,M_{\bh}\,m_{\rm H}/\left(\effrad\,\sigma_{\rm T}\,c\right)$, where $\sigma_{\rm T}$ is the Thomson scattering cross-section, and $\effrad$ is the radiative efficiency of the accretion process. To summarize, the mass-flow rate towards the black hole is $\dot{M}_{\rm accr}=\min\{\dot{M}_{\rm B},\dot{M}_{\rm Edd}\}$ and the effective black hole mass-growth rate $\dot{M}_{\bh}=\left(1-\effrad\right)\,\dot{M}_{\rm accr}$, \NEW{i.e. $\dot{M}_{\rm accr}$ here denotes the mass flow rate towards the BH through, e.g., an unresolved accretion disc}. Over a time-step $\Delta t$, the black hole mass grows by $\dot{M}_{\bh}\,\Delta t$ and a mass $\dot{M}_{\rm accr}\,\Delta t$ is removed from the gas in the cells within $r_{\rm accr}$ from the sink particle. The amount of mass removed from each cell follows the same volume-weighted scheme as the one used to compute average physical quantities in the vicinity of the sink particle.

From the disc mass-flow (accretion) rate, $\dot{M}_{\rm accr}$, we can estimate the bolometric luminosity produced in the accretion process as follows:
\begin{equation}
    L_{\rm bol} \,=\, \effrad\,\dot{M}_{\rm accr}\,c^2
    ~.\label{eqn:Lbol_accr}
\end{equation}
\noindent In this work, we assume an effective radiative efficiency $\effrad=0.1$ when calculating bolometric luminosities and accretion rates.

\subsection{Black hole feedback}
We include two modes of BH feedback that are effective in different regimes, characterised by their Eddington ratio $\lambda_{\rm Edd}\equiv \dot{M}_{\rm accr}/\dot{M}_{\rm Edd}$: a ‘radio mode' for low $\lambda_{\rm Edd}<0.01$ and a ‘quasar mode' for higher $\lambda_{\rm Edd}$. In the quasar mode, we isotropically inject thermal energy into the medium surrounding the BH, at a rate proportional to the BH bolometric luminosity, $\dot{E}_{\rm AGN}=\epsilon_{\rm q}\,L_{\rm bol}$, with a coupling efficiency $\epsilon_{\rm q}=0.1$ \citep[][]{Booth_and_Schaye2009,Dubois2012a}. In the radio mode, we transfer momentum to the gas around the BH, as a bipolar outflow with jet velocity of $v_{\rm jet}=10^4~\mathrm{km/s}$ \citep[with a mass-loading factor $\eta_{\rm kin}=100$ of the jet on unresolved scales; see, e.g.,][]{Omma2004,Cattaneo_and_Teyssier2007,Dubois2009,Dubois2010}. Kinetic energy and momentum are spread over a cone of aperture $\theta_{\rm cone}=90~\mathrm{deg}$, aligned with the average angular momentum of the gas around the BH.\\

\end{document}